\documentclass[12pt]{article}
\usepackage{amssymb}
\usepackage{graphicx}
\usepackage{graphics}
\usepackage{amsmath}

\def\l{\left(}
\def\r{\right)}
\def\la{\langle }
\def\ra{\rangle }
\newcommand{\be}{\begin{equation}}
\newcommand{\ee}{\end{equation}}
\newcommand{\ba}{\begin{align}}
\newcommand{\ea}{\end{align}}
\newcommand{\bg}{\begin{gather}}
\newcommand{\eg}{\end{gather}}
\newcommand{\bseq}{\begin{subequations}}
\newcommand{\eseq}{\end{subequations}}

\renewcommand{\Im}{\mathop{\rm Im}\nolimits}
\renewcommand{\Re}{\mathop{\rm Re}\nolimits}
\def\half{\frac{1}{2}}

\begin{document}

\title{Positronium oscillations to Mirror World revisited }

\author{S.~V.~Demidov\thanks{{\bf e-mail}: demidov@ms2.inr.ac.ru},
D.~S.~Gorbunov\thanks{{\bf e-mail}: gorby@ms2.inr.ac.ru},
\\
{\small{\em
Institute for Nuclear Research of the Russian Academy of Sciences,
}}\\
{\small{\em
60th October Anniversary prospect 7a, Moscow 117312, Russia
}}\\
A.~A.~Tokareva\thanks{{\bf e-mail}: tokareva@ms2.inr.ac.ru}
\\
{\small{\em
Faculty of Physics of Moscow State University,
}}\\
{\small{\em
Leninskiye gory 1-2, MSU, Moscow 119991, Russia
}}
}
\date{}

\maketitle

\begin{abstract}
We present a calculation of the branching ratio of orthopositronium
decay into an invisible mode, which is done in the context of Mirror
World models, where ordinary positronium can disappear from our world
due to oscillation into its mirror twin. In this revision we clarify
some formulas and approximations used  previously, correct them at
some places, add new effects relevant for a feasible experiment and
finally perform a combined analysis. We include into consideration
various effects due to external magnetic and electric fields,
collisions with cavity walls and scattering off gas atoms in the
cavity. Oscillations of the Rydberg positroniums are also
considered. 
To perform a numerical estimates in a realistic case we wrote
computer code, which can be adopted in
any experimental setup. Its work is illustrated with an example of
a planned positronium experiment within the AEgIS project. 

\end{abstract}

\section{Introduction}

Since discovery of parity violation in particle physics 
in 1957~\cite{Wu:1957my} its origin in Nature remains a mystery. An
idea for how to retain the mirror symmetry based on new particles of
opposite chiralities was put forward long ago \cite{LeeYang} and
developed to the idea of two coexisting parallel Worlds in
\cite{Okun:1966} (see also~\cite{ciar4}), where mirror transformation
can be realized as the composition of CP and
particle$\leftrightarrow$mirror particle transformations. 

The idea of a Mirror World has a long and interesting history nicely
recounted in the review \cite{Okun:2006eb}.  Today the Mirror World
model is one of the most attractive possible extensions of the
Standard Model of particle physics.  This theory is based on
the assumption that in addition to the Standard Model its mirror image exists 
and corresponding new particles interact very weakly with SM
fields. Thus, the theory is based on the gauge group $G\times G'$,
where $G=SU(3)\times SU(2)_L\times U(1)_{Y}$, $G'=SU(3)'\times
SU(2)_R'\times U(1)^{\prime}_{Y}$. As it was pointed out before, originally the
deep theoretical motivation for this extension was originally a
restoration of the 
mirror symmetry: P-transformation changes ordinary left particles to
mirror right ones~\cite{LeeYang}. If this symmetry is exact then every
particle has its mirror partner (twin) with the same mass and lifetime but
is charged under different gauge groups. 

Certainly, if mirror particles exist they inevitably interact with
ordinary particles gravitationally. But in general these models allow for
two kinds of {\it direct} renormalizable interactions --- portals --- between the
Standard Model (SM) and the Mirror World particles. One of them,
called the Higgs portal~\cite{higgs1}, can be probed in high energy
collider experiments (see {\it e.g.} Ref.~\cite{higgs2}). Here we
discuss implications of the second type of interactions between SM and
Mirror World particles, dubbed as the Abelian portal,  
\begin{equation}
\label{U1-portal}
  {\cal L}_Y =
  \frac{\epsilon}{2\cos{\theta_{W}}\cos{\theta_{W}^{\prime}}}
  B'_{\mu\nu}B^{\mu\nu}\;, 
\end{equation} 
where $B_{\mu\nu}$ and $\theta_{W}$ are the $U(1)_Y$ field strength
tensor and the weak mixing angle, respectively, and $B_{\mu\nu}^{\prime}$
and $\theta_{W}^{\prime}$ are the same quantities in the mirror
sector. Studying of this interaction calls for another sort of
experiment. 

The most sensitive way to search for the Mirror World in a laboratory is
looking for the disappearance of ordinary orthopositronium (oPs) resulting
from oscillation to the mirror twin and its subsequent decay to
mirror photons. The oscillation becomes possible due to a photon - mirror
photon mixing hidden in Eq.~\eqref{U1-portal}. Similar processes for
bound states of quarks (vector mesons) are much less sensitive to the Abelian portal~\eqref{U1-portal}.

A new experiment~\cite{Crivelli:2010bk} is proposed to search for the
oPs disappearance with the sensitivity to branching ratio at the
level of $10^{-7}-10^{-8}$. This refers to the value of the mixing
parameter entering~\eqref{U1-portal} $\epsilon\sim 10^{-9}$. Quite
remarkebly, it is the first time when experimentalists have a chance
to enter a region of parameter space which is both phenomenologically
viable and interesting for cosmological applications.

Indeed, the mirror matter can play very important role in cosmology
and astrophysics (see, {\it e.g.} Refs~\cite{Blinnikov:1983gh,
  Berezhiani:2008gi, Ciarcelluti:2010zz}). Mirror matter can behave 
like dark matter on the astrophysical and cosmological scales. With
Mirror symmetry (slightly) violated (say, by different vacuum
expectation values of our Higgs field and Mirror Higgs field) the
phenomenology 
becomes more fascinating. Indeed, photon-paraphoton mixing from the
Abelian portal~\eqref{U1-portal} implies the presence of millicharged
particles: mirror particles effectively carry a tiny charge with respect 
to electromagnetism. Thus, mirror matter particles can be directly
produced in collider experiments~\cite{LHC}, be tested with
electromagnetic precision measurements~\cite{probe of millicharge}, 
change the CMB anisotropy picture due to participation in primordial plasma
dynamics at recombination~\cite{Dubovsky:2003yn}, contribute to
supernovae explosions~\cite{Raffelt}. 
The idea of Mirror World naturally explains the presence of
light sterile neutrino(s) at the eV-scale, as suggested by neutrino
anomaly announced by the LSND experiment~\cite{LSND} and is now in
agreement with the 
combined analysis of cosmological data~\cite{sterile neu 1,sterile neu
   2,sterile neutrino 3}.
Indeed, since we have sub-eV neutrinos in our sector, which within the
SM acquire masses from dimension-5 interaction terms, then mirror 
(sterile)
neutrino masses are of the same order. Mirror symmetry provides 
similar dimension-5 operators which give masses to mirror neutrinos
and may yield mixing with neutrinos from our sector.

In this paper we revise positronium oscillations into its mirror
twin. The paper is organized as follows. We start in
Section~\ref{Sec:Warming} with the discussion of positronium oscillations
in a vacuum. Further, in Section~\ref{section3} we describe the relevant
parameters of the proposed experiment~\cite{Crivelli:2010bk}
(Section~\ref{AEgIS}), present the combined analysis of positronium 
scattering off the gas atoms 
(Section~\ref{Sec:gas}) and the cavity walls
(Section~\ref{Sec:walls}), and take into account the interaction with electric
and magnetic fields (Section~\ref{Sec:fields}). Numerical estimates are
given for a particular setup of
the proposed experiment~\cite{Crivelli:2010bk}. Section~\ref{Sec:Rydberg} is
devoted to the study of oscillations of the Rydberg (highly excited)
positroniums.
Section~\ref{Sec:Conclusions} contains conclusions and a
discussion of open problems. 

\section{Warming-up task: orthopositronium oscillations in vacuum}
\label{Sec:Warming}

We begin with a simplified picture of vacuum oscillations.  
Hereafter if not stated otherwise we consider a model
with the Mirror World, where the mirror symmetry is exact: values of all
mirror masses and mirror coupling constants coincide with those of
the corresponding originals. Positronium couples to
its  mirror twin due to photon-paraphoton mixing emerging from the
Abelian $U(1)_Y$-portal 
\eqref{U1-portal}, 
\begin{equation}
\label{photon-mixing}
 {\cal L}_{int}=\frac{\epsilon}{2}\, F'_{\mu\nu}F^{\mu\nu}\;,
\end{equation}  
where $F_{\mu\nu}$ and $F^{'}_{\mu\nu}$ are electromagnetic and
paraelectromagnetic field strength 
tensors, respectively. Orthopositronium (oPs) can oscillate
to its twin oPs$'$ via virtual photon mixed with virtual
paraphoton due to coupling \eqref{photon-mixing}. 
The corresponding transition matrix element 
reads 
\begin{equation}
\label{positronium-mixing}
\la \mathrm{oPs}\mid {\cal L}_{int}\mid \mathrm{oPs}'\ra
=2\pi\epsilon f\equiv\frac{\delta}{2}\;,
\end{equation} 
where parameter 
\[
f\approx8.7\times10^4\;{\rm MHz}
\]
is determined \cite{Glashow:1985ud} 
by the one-photon annihilation diagram involving orthopositronium.
Experimental constraints on mixing parameter
$\epsilon$ have been recently reviewed in Ref.~\cite{Berezhiani:2008gi}. 
The first experimental limits on $\epsilon$ from oPs physics were
discussed in Refs.~\cite{Gninenko:1994dr,Foot:2000aj}. Searches for an
invisible decay of orthopositronium have a rather long
history~\cite{Atoian:1989tz,Mitsui:1993ha,Gninenko:2006sz}.  Nonobservation of 
positronium disappearance  $\mathrm{oPs}\rightarrow invisible$ places 
a direct limit on the mixing parameter~\cite{Badertscher:2006fm} 
\[
\epsilon<1.55\times10^{-7}\;. 
\] 

The strongest indirect limit on $\epsilon$ in a model with
exact mirror symmetry comes from cosmology~\cite{Berezhiani:2008gi},  
\begin{equation}
\label{cosmo-limit}
\epsilon<3\times10^{-10}\;.
\end{equation}
It ensures that our and Mirror Worlds have never come to equilibrium
in the early Universe; then with somewhat colder mirror plasma the
mirror baryons serve as cold dark matter in the Universe. Remarkably,
the limit \eqref{cosmo-limit} is not far from the
estimate~\cite{Foot:2011pi}   
\begin{equation}
\label{dama-expl}
\epsilon\sqrt{\xi_{O'}}\approx \l 7\pm3\r\times10^{-10}\;,
\end{equation}
required to explain the annual modulation signal, in
direct dark matter searches performed by CoGeNT and DAMA experiments.
Here $\xi_{O'}$ is a mass fraction of the 
mirror oxygen in dark matter particles. 
Strictly speaking, there is no overlap between the two ranges
\eqref{dama-expl} and \eqref{cosmo-limit}. The gap grows
on account of the fact that observation of the matter
distribution in the Bullet cluster (1E 0657-558)
can be reconciled with the Mirror
dark matter only if the largest fraction of dark matter particles are
confined in compact macroscopic objects (mirror asteroids, stars, etc)
\cite{Blinnikov:2009nn}. Then, the dark matter particle flux should be
lower than what was supposed in \cite{Foot:2011pi} and hence the
required DAMA/CoGeNT signal interval of the mixing parameter is higher
than \eqref{dama-expl}. At the same time, the analyses of new data on 
primordial Nucleosynthesis \cite{arXiv:1001.4440} and cosmic microwave
background anisotropy \cite{arXiv:1001.4538} suggest a somewhat higher
rate of the Universe expansion which can be explained by a contribution of
mirror matter. This relaxes the cosmological limit \eqref{cosmo-limit}
to some extent. 

In the nonsymmetric model, the cosmological bound is generally much
weaker. For instance, let the only source of asymmetry be different
values of the Higgs boson vacuum expectation values, $\langle H \rangle
\neq \langle H' \rangle$. Then in the hierarchical case $\langle H
\rangle \ll \langle H' \rangle$ one obtains~\cite{Berezhiani:2008gi}
\[
\epsilon<3\times10^{-9}\times\sqrt{\frac{\langle H' \rangle}{\langle
    H\rangle}}\;, 
\]
from the succesfulness of the Big Bang Nucleosynthesis. 
 
To summarize the limits discussed above, in what follows we adopt 
\begin{equation}
\label{reference-mixing} 
\epsilon=1\times 10^{-9}
\end{equation}
as a reference number for numerical estimates. This particular value
is of considerable interest, since it has been argued
\cite{Foot:2008nw} that the DAMA/LIBRA periodic
signal\,\cite{Bernabei:2008yi} and observations by  
CoGeNT\,\cite{Aalseth:2010vx} 
can be explained within the Mirror World
model with a mixing of the order~\eqref{reference-mixing} by galactic mirror
baryons, while other relevant experiments~\cite{Angle:2007uj} 
were blind to this dark matter. In any case, if mirror matter plays the
role of dark matter particles, the direct limit on mirror mixing
should be of order~\eqref{reference-mixing}.   
Moreover, it has been argued that studies of positronium in the
proposed experiment~\cite{Crivelli:2010bk} will be sensitive right up
to the values of mixing parameter. Thus, for the first time the 
experiment enters into the phenomenologically interesting and
cosmologically allowed region of the model parameter space. 

For the reference mixing~\eqref{reference-mixing} the value of matrix
element responsible for $\mathrm{oPs}\leftrightarrow \mathrm{oPs}'$ 
oscillations~\eqref{positronium-mixing} is  
\begin{equation}
\label{reference-oscillation-rate}
\delta\approx 1090\,{\rm s}^{-1}\approx 7.3\times10^{-13}\,{\rm eV}\;,  
\end{equation}  
which is much lower than the rate of orthopositronium decay (mostly
decay to three photons) \cite{637617,hep-ex/0308030} (see
Ref~.\cite{arXiv:0806.4927} for more precise and recent theoretical
calculations),
\begin{equation}
\label{positronium-lifetime}
 \gamma\approx 7.040\times10^6\,{\rm s}^{-1}\approx 4.634\times 10^{-9}\,{\rm eV}\;.
\end{equation}

Hereafter {\em we work in the ``flavor'' basis:} 
the evolution of the system oPs-oPs$'$ in a vacuum can be described by
the Schr\"odinger equation for the doublet wave function $\Psi=\l
\psi_{\mathrm{oPs}}\,, \psi_{\mathrm{oPs}'}\r^{\rm T}$, 
\begin{equation}
\label{dy}
i\frac{d\Psi}{dt}=H\Psi\;,
\end{equation}
where the Hamiltonian in the positronium rest frame accounts for
the positronium and mirror positronium decays and their mutual
oscillations,  
\begin{equation}
\label{vacuum-hamiltonian}
H=\l
    \begin{array}{cc}
     E-i\gamma/2  &   \delta/2 \\
      \delta/2 &   E -i\gamma/2
          \end{array}
    \r\;,
\end{equation} 
where $E$ is the kinetic energy of oPs and oPs$'$. 
When solving Equation \eqref{dy} for the wave function one obtains the
probability to have mirror positronium instead of initial positronium 
by the elapsed time $t$,   
\begin{equation}
P\l \mathrm{oPs}\rightarrow \mathrm{oPs}'\r=e^{-\gamma t}\,\sin^2
\frac{t\delta}{2}
\;.
\end{equation}
Mirror orthopositronium decays mainly into three mirror photons and 
the probability of oPs disappearance is given by 
\cite{Feinberg:1961zza} 
\begin{equation}\label{branching-in-vacuum}
  {\rm Br}\l \mathrm{oPs}\to \mathrm{invisible}\r = 
\gamma \int^{\infty}_0 P\l \mathrm{oPs}\rightarrow \mathrm{oPs}'\r dt=\frac{\delta^2}{2}
  \frac{1}{\gamma^2
    +\delta^2}\approx \frac{\delta^2}{2\gamma^2}\;,
\end{equation}
since $\delta\ll \gamma$, cf. Eqs.~\eqref{reference-oscillation-rate} 
and \eqref{positronium-lifetime}. 

\section{Realistic consideration}
\label{section3}

\subsection{Realistic experimental setup}
\label{AEgIS}

We proceed with the study of complications in the positronium description
arisen in a real experiment. Indeed, the realistic case of positronium
oscillations is much more involved. In our analysis we consider the
general setup of experiments proposed in
Ref.~\cite{Badertscher:2003rk} and then further developed in
Ref.~\cite{Crivelli:2010bk} with more reliable estimates of the
sensitivity to the mixing parameter.  

In the experiment~\cite{Crivelli:2010bk} oPs states will be formed
in a thin nanoporous SiO$_2$ target placed on the bottom of 
the vacuum cavity (10
cm in diameter and 10 cm in height). After the formation a significant
fraction of the 
oPs can  escape inside the cavity and can become almost completely
thermalized  with the average temperature equal to the temperature of 
the target~ \cite{637617}. The cavity is surrounded by an almost
hermetic $4\pi$ electromagnetic calorimeter to detect annihilation
photons. The experimental signature of the oPs-oPs$'$ oscillations is
the  absence of energy deposition in the calorimeter expected from
the ordinary positronium decays. The typical residual gas pressure is
above $10^{-5}$\,Torr. There is an optional external electric field $E\lesssim
100$\,kV/cm and magnetic field $B\lesssim 10^3$\,G (exceptionally
up to $B\lesssim 10^5$\,G). 

Note that the experiment~\cite{Crivelli:2010bk} can also be
preformed, {\it e.g.}, in the framework of the AEgIS project
(Antihydrogen Experiment: Gravity, Interferometry, Spectroscopy) at
CERN~\cite{Doser:2010zz,aegis_site}. The main goal of AEgIS experiment 
is a mass production of 
antihydrogen. In this experiment orthopositroniums are a source of
slow positrons that recombined with antiprotons. There is a proposed
technology of 
adopting Rydberg (highly excited) positroniums instead of the usual
ground-level orthopositroniums: the former live longer and positroniums move
slower inside. Thus, in the AEgIS experiment, not only ground states
will by confined within the cavity,
but also excited oPs with the Rydberg numbers $n\lesssim 30$. 

\subsection{Oscillations in gas}
\label{Sec:gas}

Let us consider the role of positronium interaction with
gas. Generally, the time-evolution of the system oPs-oPs$'$ is
described by density matrix
\begin{equation}
\label{density-matrix}
\rho \l t\r=\int\!\! d^3x~ \Psi \Psi^\dagger =\int \!\! d^3x
\l
    \begin{array}{cc}
     \psi_{\mathrm{oPs}}\psi_{\mathrm{oPs}}^*  &   \psi_{\mathrm{oPs}}^*\psi_{\mathrm{oPs}'} \\
      \psi_{\mathrm{oPs}}\psi_{\mathrm{oPs}'}^* &   \psi_{\mathrm{oPs}'}\psi_{\mathrm{oPs}'}^*
          \end{array}
    \r\;.
\end{equation}
The density matrix \eqref{density-matrix} solves the following
equation \cite{Feinberg:1961zza} 
\begin{equation}\label{1}
\frac{d\rho}{dt} = -i {\cal H} \rho + i\rho {\cal H}^\dagger 
+ 2\pi\,n\,v\!\int \!\! d\cos\theta \;F\!\l
\theta\r \rho\,
  F^\dagger\!\!\l\theta\r\;,
\end{equation}
with the Hamiltonian, cf. \eqref{vacuum-hamiltonian}, 
\begin{equation}
\label{hamiltonian-in-gas}
    {\cal H}  = \l
    \begin{array}{cc}
      -\frac{2\pi}{k}\, n\, v\, f\!\l 0\r+E-i\,\gamma/2  &   \delta/2 \\
      \delta/2                               &   E^{\prime}-i\,\gamma^{\prime}/2
          \end{array}
    \r\;,
\end{equation}
where $E$ and $E^{\prime}$ are the positronium and  mirror positronium
energies, $\gamma$ and $\gamma^{\prime}$ are their widths\footnote{
Different notations for oPs and oPs$^{\prime}$ are used
($E$, $E^{\prime}$ for energies and $\gamma, \gamma^{\prime}$ for
widths) because corresponding quantities can differ from one another
in external
electromagnetic field and for the case of broken mirror symmetry.
}, $k$ refers
to the value of the positronium 3-momentum, and $v$ is the mean
relative velocity 
between the positronium and gas molecules. The positronium scattering off
gas is described by the matrix
\begin{equation}
\label{amplitude}
    F\l\theta\r=\l
    \begin{array}{cc}
     f\!\l\theta\r  &   0 \\
      0         &   0
          \end{array}
    \r\;,
\end{equation}
where $\theta$ is a scattering angle; in the cavity 
{\it mirror positronium } does not
scatter off {\it ordinary matter } given the small mixing 
\eqref{reference-mixing}. The first term 
in the Hamiltonian \eqref{hamiltonian-in-gas} accounts for
the presence of gas in the system, which changes the energy 
of the positronium state. There $n$ stands for the 
number density of gas particles, so that the frequency of the positronium
scatterings off gas $w$ is determined by the cross section $\sigma$ or
(with help of the optical theorem) by the forward scattering amplitude
$f\!\l 0\r$ as 
\begin{equation}\label{imagine-amplitude}
w\equiv\sigma\, n\, v=\frac{4\pi}{k}\, n\, v\, \Im\! f\!\!\l 0\r\;.
\end{equation}

In what follows, we resort to the case of slow scatterings, where the
scattering amplitude does not actually depend on the scattering angle
$\theta$: {\it i.e.} scattering is saturated in $s$-wave. The real part of
the scattering amplitude is positive\footnote{At large distances $r$
  positronium is attracted by an atom, {\it i.e.}, it propagates in the
  negative potential $U\l r \r<0$, hence in the Born approximation for
  nonrelativistic particles we obtain positive value for the amplitude
  (see, {\it e.g.} \cite{landau3}), $f\approx
  -2m\int_0^{\infty}U(r)r^2dr>0$. }   
and by analogy with Eq.~\eqref{imagine-amplitude}, 
can be conveniently parametrized as  
\begin{equation}\label{real-amplitude}
w_{Re}\equiv \frac{4\pi}{k}\, n\, v\, \Re\! f\!\!\l 0\r\;.
\end{equation}
In this case the problem is solved analytically as follows. Once
positronium is produced (at $t=0$), the system is in the pure flavor
state,
\begin{equation}
\label{initial-state-with-gas}
   \Psi\l 0\r= \l
    \begin{array}{c}
     1  \\
     0  
          \end{array}
    \r\;,
~~~~~~
 {\rm and~hence}~~~~~~\rho\l 0\r = \l
    \begin{array}{cc}
     1  &   0 \\
     0  &   0
          \end{array}
    \r\;.
\end{equation}
We parametrize the density matrix at arbitrary moment as 
\be
    \rho\equiv\l
    \begin{array}{cc}
     \rho_1       &   x+iy \\
     x-iy    &  \rho_2
          \end{array}
    \r\;,
\ee
then Eq.\,\eqref{1} yields 
\begin{subequations}
\label{equations-with-gas}
\begin{align}
   \frac{d\rho_1}{dt}&=-\gamma \rho_1-\delta y \\
   \frac{d\rho_2}{dt}&=\delta y-\gamma^{\prime} \rho_2 \\
 \frac{dy}{dt} & = \frac{\delta}{2}
 \l \rho_1-\rho_2\r +\l \frac{w_{Re}}{2}-\Delta\r x -
 \frac{1}{2}\l \gamma+\gamma^{\prime}+w\r y 
\label{equations-with-gas-3}
\\  
  \frac{dx}{dt} & = -\frac{1}{2}\l \gamma + \gamma^{\prime}
  +w\r x+\l \Delta-\frac{w_{Re}}{2}\r y 
\label{equations-with-gas-4}
\end{align}
\end{subequations}
where $\Delta\equiv E^{\prime}-E$. 

We are interested in the total probability for the orthopositronium to
disappear, which under study reads 
\be 
\label{add-label-1}
{\rm Br}\l \mathrm{oPs}\to
\mathrm{invisible}\r=\frac{\int_0^{\infty}\rho_2 dt }{\int_0^{\infty}\rho_1 dt
  +\int_0^{\infty}\rho_2 dt}\;. 
\ee
One can integrate the system \eqref{equations-with-gas}, with the initial
condition \eqref{initial-state-with-gas}, and introducing notations 
\[
Y\equiv \int_0^{\infty}y dt\;,~~~~X\equiv \int_0^{\infty}x dt\;,~~~~
P_i\equiv \int_0^{\infty}\rho_i dt\;,~~i=1,2\;,
\]
arrive at a linear system of algebraic equations:  
\begin{equation}
\label{equations-with-gas-linear}
\begin{split}
   -\gamma P_1-\delta Y&=-1 \\
   \delta Y-\gamma^{\prime} P_2&=0 \\
 \frac{\delta}{2} \l P_1-P_2\r +\l \frac{w_{Re}}{2}-\Delta\r X -  
\frac{1}{2}\l\gamma+\gamma^{\prime}+w\r Y&=0  \\  
  -\frac{1}{2}\l\gamma + \gamma^{\prime}+w\r
  X+\l\Delta-\frac{w_{Re}}{2}\r Y&=0 
\end{split}
\end{equation}
Then the oscillation probability (or probability to disappear)
\eqref{add-label-1}  
is given by
\be
{\rm Br}\l \mathrm{oPs}\to
\mathrm{invisible}\r = \frac{P_{2}}{P_1 + P_2}\;.
\ee
Let us introduce the following notations:
\begin{equation}
\label{intro-parameters}
\Gamma\equiv \frac{1}{2}\l\gamma+\gamma^{\prime}+w\r\;,~~~~~~~~
\Delta_{Re}\equiv\Delta-w_{Re}/2\;.
\end{equation}
Taking into account the hierarchy
$\delta\ll\Gamma$, the solution of the
system~\eqref{equations-with-gas-linear} can be simplified and we obtain $P_1
+ P_2 = \frac{1}{\gamma}\l 1 + {\cal O}\l
\frac{\delta^3}{\Gamma^3}\r\r$ and for the
probability~\cite{Feinberg:1961zza},
\be \label{main}
{\rm Br}\l \mathrm{oPs}\to
\mathrm{invisible}\r = \frac{\delta^2}{2}
\frac{\Gamma}{\gamma^{\prime}}\frac{1}{\Gamma^2 + \Delta_{Re}^2} \l 1+
     {\cal O}\l\frac{\delta^2}{\Gamma^2}\r\r \;.
\ee
Note that in the ``flavor'' basis, the case of an exact 
mirror symmetry implies equal energies and widths of positronium and
its twin in the absence of external electromagnetic fields, hence
$\Delta =0$ and $\gamma^{\prime} = \gamma$. 

For the numerical estimate, we 
apply the obtained formula to a particular case of the experiment
described in Section\,\ref{AEgIS}. Here the cavity is proposed to be
filled with nitrogen $N_2$ at room temperature $T_0$, so that
\begin{equation}\label{gas-temperature}
k_BT_{0}=0.025\,{\rm eV}\;.
\end{equation}

The momentum transfer cross section of positronium on nitrogen was
  measured in Ref.~\cite{cross} as follows:
\be
\label{nitrogen-cross-section}
\sigma_{m}\equiv\int\l 1-\cos{\theta}\r\frac{d\sigma}{d\Omega}d\Omega
= \l 35\pm 8\r\times 10^{-16} {\rm cm}^2\;,
\ee
and remains almost constant with temperature varying at
$k_BT_{0}<0.3$\,eV. And if scattering amplitude does not depend on 
scattering angle we obtain for the total cross section after
integrating over the angle variables $\sigma\equiv\int
\frac{d\sigma}{d\Omega}d\Omega \approx \sigma_{m}$. 
In accordance with the feasible experimental setup~\cite{Crivelli:2010bk} we
assume that orthopositroniums are emitted from the target with a
temperature of $T_{\mathrm{oPs}}\sim T_0$, close to the nitrogen
temperature \eqref{gas-temperature}. Thus, the positronium average
momentum is $k=m_{\mathrm{oPs}} v_{\mathrm{oPs}}$ where positronium
average velocity equals 
\begin{equation}
\label{positronium-velocity}
v_{\mathrm{oPs}}=\sqrt{\frac{3k_B
    T_{\mathrm{oPs}}}{m_{\mathrm{oPs}}}}=8.1\times 10^6\,{\rm
  cm}\,{\rm s}^{-1}\times \sqrt{\frac{k_B
    T_{\mathrm{oPs}}}{0.025\,{\rm eV}}}\;. 
\end{equation}
In principle, there may be processes of pick-off annihilation and
ortho-to-parapositronium conversion. But the corresponding probabilities are
negligibly small (see, {\it e.g.} Ref.~\cite{goldansky}) at the positronium
energies and the gas pressures relevant for the class of experiments in
question~\cite{Crivelli:2010bk}. 

In what follows, we set $T_{\mathrm{oPs}}=T_{0}$. One observes, that
$v_{\mathrm{oPs}}$ is much higher than the velocity of nitrogen atoms, so
one can replace $v$ in Eq.~\eqref{imagine-amplitude} with
$v_{\mathrm{oPs}}$ in order to estimate the collision frequency and,
in particular, to check that for the cross section
\eqref{nitrogen-cross-section}, the slow scattering approximation is 
valid indeed and the scattering amplitude does not depend on the
scattering angle as we have assumed. For corresponding energies of
colliding particles the cross section is mostly saturated by the real
part of the amplitude, so 
\[
\left| \Re f\!\l 0\r\right|\approx \sqrt{\sigma/4\pi}\;.
\] 

The nitrogen pressure $P_{N_2}$ is related to the
nitrogen number density $n_{N_2}$ by $P_{N_2}=n_{N_2}k_B
T_{N_2}$. Obviously, the presence of gas  suppresses oscillations
(see Eq.~\eqref{main}), if its
density is high 
enough for positronium to scatter off atoms at least once. 
This effect can be understood as the loss of coherence due to
a different angular distribution of the positronium and its mirror twin
after collision, which is encoded in Eq.~\eqref{amplitude} (see also
Ref.~\cite{MPI-PAE/PTh 54/80} for discussion of similar effects). 
For the
reference temperature~\eqref{gas-temperature}  this implies a pressure
$P_{N_2}$ of order $10^{-3}$\,Torr or higher. We put all the formulas above
into Eq.~\eqref{main} with the reference value of the mixing
parameter~\eqref{reference-mixing}. When introducing dimensionless
variables  ${\cal P}\equiv P_{N_2}/\left(10^{-3}\right.$\,Torr$\left.\right)$ and ${\cal T}\equiv 
k_B\,T_{N_2}/\left(0.025\right.$\,eV$\left.\right)$ we write for the invisible branching ratio 
\be \label{main-in-numbers}
{\rm Br}\l \mathrm{oPs}\to
\mathrm{invisible}\r=1.2\times 10^{-8}\times \l \frac{\epsilon}{10^{-9}}\r^2 
\times \frac{1+0.067\times {\cal P}/{\cal T}^{1/2}}{(1+0.067\times{\cal
    P}/{\cal T}^{1/2})^2 + (0.29\times{\cal P}/{\cal T})^2}\;.
\ee
This dependence can be used~\cite{Gninenko:1994dr} to check
against the possible systematics once the signal is found. 

\subsection{Oscillations in a finite volume}
\label{Sec:walls}

In a laboratory orthopositronium is trapped in a cavity. Then
if positronium velocities are high enough to reach the walls before
decay, the interaction with wall material modifies the (possible)
oscillations into mirror positronium. The reason is simple: positronium
interacts with walls, while its twin does not and if produced in a
cavity flies away freely. 

Let us use the instant approximation for the positronium interaction
with a wall. Then, between the scatterings off the wall, the
oscillations proceed as in an infinite volume case, but when positronium
hits the wall, nondiagonal elements of the density matrix
\eqref{density-matrix} nullify. The oscillations stop at this 
point, since the mirror positronium flies away, while the positronium gets
reflected, and hence their wave functions get separated \emph{in space},
and the coherence gets lost. This is the main observation.      

To account for this instant decoherence properly, let us consider the
oscillations of a stable positronium. Then, the evolution of the
density matrix elements is described by the following equations:
\begin{subequations}
\label{system}
\begin{align}
\frac{d\rho_1}{dt}&=-\delta y 
\\
\frac{d\rho_2}{dt}&=\delta y 
\\
\label{y-equation}
\frac{dy}{dt}&=\frac{\delta}{2}(\rho_1-\rho_2)
\end{align}
\end{subequations}
and $x$ remains zero with initial condition
\eqref{initial-state-with-gas}, which is the pure positronium state. 

It is convenient to introduce the variable $s\equiv \rho_1-\rho_2$ and
parametrize the time interval between $i\!-\!1$-th and $i$-th 
reflections as $\tau_i$; the system starts to evolve at $t=0$ being in
the positronium state \eqref{initial-state-with-gas}. Then before the first
reflection the solution of the system~\eqref{system} is 
\[
t<\tau_1\;:~~~~s=\cos t \delta\;,~~~y=\half\, \sin t \delta\;.
\]
At $t=\tau_1$ we nullify $y\l \tau_1 \r=0$, but the diagonal variable
remains intact, $s\l \tau_1\r=\cos \tau_1 \delta$. Then, the
evolution of the system between the first and second reflections
proceeds as follows,  
\[
\tau_1<t<\tau_1+\tau_2\;:~~~~s=\cos\tau_1\delta \, \cos\l t-\tau_1\r\!
\delta\;,~~~ y=\half \,\cos\tau_1\delta \,\sin \l t-\tau_1\r\!\delta\;.
\] 
It is straightforward to see that right after the $n$-th reflection the solution
reads 
\[
s\l \tau_n\r =\prod_{k=1}^n \cos \tau_k \delta \;,~~~ y \l \tau_n\r=0\;. 
\] 

When introducing the average time between reflections 
\[
\la \tau \ra \equiv \frac{t}{n}
\]
one finds that the positronium velocity
\eqref{positronium-velocity} and the typical size $L\sim 10$\,cm 
of the detector volume are related as $\la \tau \ra \simeq
L/v_{\mathrm{oPs}}$. For the reference value of the oscillation
rate \eqref{reference-oscillation-rate}, the average reflection rate
obeys 
\[
\la \tau \ra \delta \ll 1\;.
\]
Then we can use the exponential approximation to obtain the smooth
solution 
\be \label{2}
s\approx
\exp\l \sum_{k=1}^{n}\log\l 1-\frac{\l \tau_{k}\delta\r^2}{2}\r\r
\approx
\exp\l-\frac{\delta^2n}{2}\frac{1}{n}\sum_{k=1}^{n}\tau_{k}^{2}\r =
\exp\l-\frac{\delta^2\langle\tau^{2}\rangle}{2\langle\tau\rangle}t\r  
\ee
Thus, collisions with walls result in the exponential suppression of
oscillations. 

One can mimic the suppression \eqref{2} by introducing an additional
suppression to the Equation \eqref{y-equation}, 
\begin{equation}
\label{mimic-10}
\frac{dy}{dt}=\frac{\delta}{2}s-{\rm w}y\;.
\end{equation}
Then, at $\delta/{\rm w}\ll 1$ one has 
\[
s=\exp\l-\frac{\delta^2t}{\rm w}\r\;,
\]
and matching \eqref{2} is achieved with 
 \be
\label{twice}
 {\rm w} =\frac
{2\la\tau\ra}{\la\tau^2\ra}\;.
\ee
Thus, comparing \eqref{mimic-10} to \eqref{equations-with-gas-3} one
accounts for the collisions with walls by replacing 
(see \eqref{intro-parameters}) 
\[
\Gamma\to \Gamma+{\rm w}
\]
in the formula for oscillations \eqref{main}.  

To support our approach, let us consider a limiting case when the time
intervals between the collisions with the walls are all equal,
$\tau_{k+1}-\tau_k\equiv\tau$, then $\la\tau^2\ra^{1/2} = \la\tau\ra 
 = \tau$ and one obtains for the oscillation probability 
\be
\rho_2 = \frac{1}{2}\l 1 - {\rm e}^{-\frac{\delta^2\,\tau t}{2}}\r
 \approx \frac{1}{4}\delta^2\,\tau t.
\ee
This result is in agreement with
Refs.~\cite{Kerbikov:2002uq,Kerbikov:2008zz} for the neutron-antineutron
and neutron-mirror neutron systems where any neutron collides for many
times. On the contrary, the obtained result deviates
 from a similar one in Ref.~\cite{Crivelli:2010bk} where ${\rm w}=1/\la 
 \tau \ra$ instead of \eqref{twice}, hence it is two times smaller. 
The origin of this  
term in the corresponding equations of Ref.~\cite{Crivelli:2010bk} was
not explained in that paper. 
 One can obtain this answer with our method provided the random
 variable $\tau$ is distributed in accordance with the Poisson statistics,
 which we find unjustified when the positronium trajectories are reasonably
 tractable.  

However, in general case, the wall
collision rate is not a quantity which can be well-defined in any real
experiment. For example, 
the authors of Ref.~\cite{Crivelli:2010bk} argue that the
number of collisions in the proposed experiment 
is of order unity, {\it i.e.} $\tau
\gamma \sim 1$. In this case, the gas pressure,
the geometry of cavity, and the properties of the oPs source (like energy
and angle distributions) become important and the wall collision rate 
becomes a largely uncertain quantity (e.g., it depends on the
direction of positronium propagation, etc.).  

To take into account all those effects and make accurate predictions 
for the probability of positronium disappearance, 
a numerical calculation should be performed. We apply the Monte-Carlo
approach to simulate the evolution of the oPs-oPs$'$ system in the
following way. The density matrix is calculated numerically by
solving Eq.~\eqref{equations-with-gas} with initial
conditions~\eqref{initial-state-with-gas}. We use the geometry of the
positronium cavity described in Section~\ref{AEgIS}. In our numerical
simulations, we suppose that the positroniums in the initial state have
isotropic nonrelativistic Maxwell velocity distribution with a
temperature of $T_{\mathrm{oPs}}$ (equal to the temperature of gas),
which is reasonably close to the realistic distribution observed in
Ref.~\cite{637617}. When collision with the wall takes place
we change the density matrix in the following way 
\begin{equation}
\rho\equiv\left(
\begin{array}{ll}
\rho_{11} & \rho_{12} \\
\rho_{21} & \rho_{22}
\end{array}
\right) \to \left(
\begin{array}{ll}
\rho_{11} & 0 \\
0 & 0
\end{array}
\right)\;,
\end{equation}
which corresponds to total disappearance of mirror positronium from
the cavity. The decoherence due to gas collisions is already taken
into account in Eq.~\eqref{equations-with-gas}. However, collisions
with gas also change the direction of positronium
motion. 
\begin{figure}[!htb]
\includegraphics[angle=0,width=0.80\columnwidth]{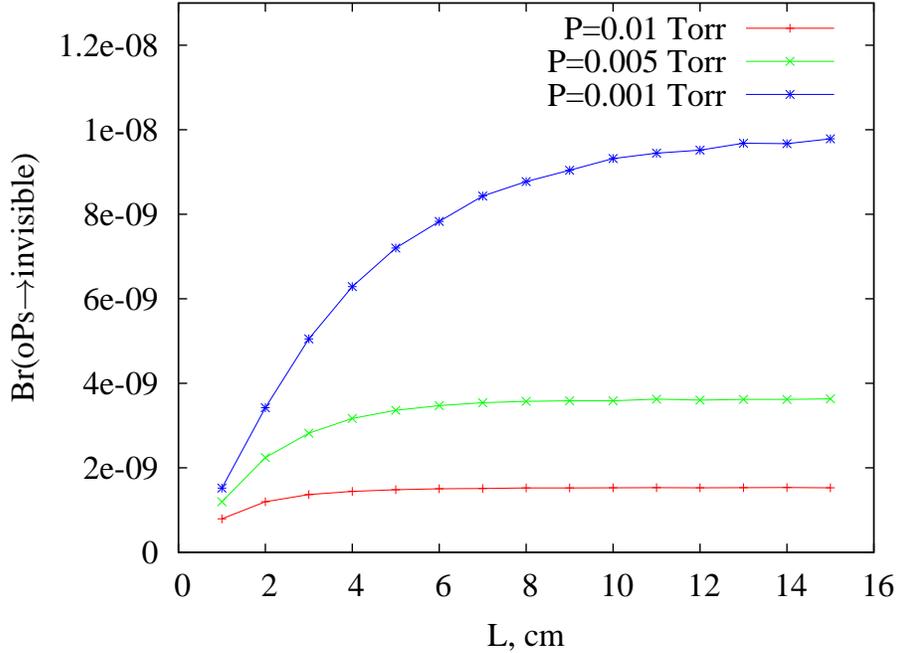} 
\caption{\label{Lx} The dependence of the branching ratio on
  the characteristic size of experiment cavity $L$ at different
  values   of gas pressure. We chose $\epsilon=10^{-9},
  k_{B}T_{\mathrm{oPs}}=0.025$\,eV. 
}
\end{figure}
So, we simulate time intervals between the collisions according to its
average value~\eqref{imagine-amplitude}: the probability of the
positronium experiencing an interaction in small time interval $dt$
equals $P_{int} = \rho_{11}\left(1-{\rm e}^{-wdt}\right)$. After
the collision with a gas atom, the direction of the positronium velocity changes
according to the isotropic distribution. We neglect small changes in the
absolute velocity in the scattering off gas atoms. To get feeling of the
influence of wall collision let us consider the same topology of
experiment cavity as it is described in Section~\ref{AEgIS} but with
different absolute size, {\it i.e.} we take vacuum cavity of $L$ cm in
diameter and $L$ cm in height. In Fig.~\ref{Lx} we present the
branching ratio dependence on the characteristic size $L$ of cavity
where orthopositroniums decay.  

\begin{figure}[!htb]
\begin{tabular}{ll}
\includegraphics[angle=0,width=0.45\columnwidth]{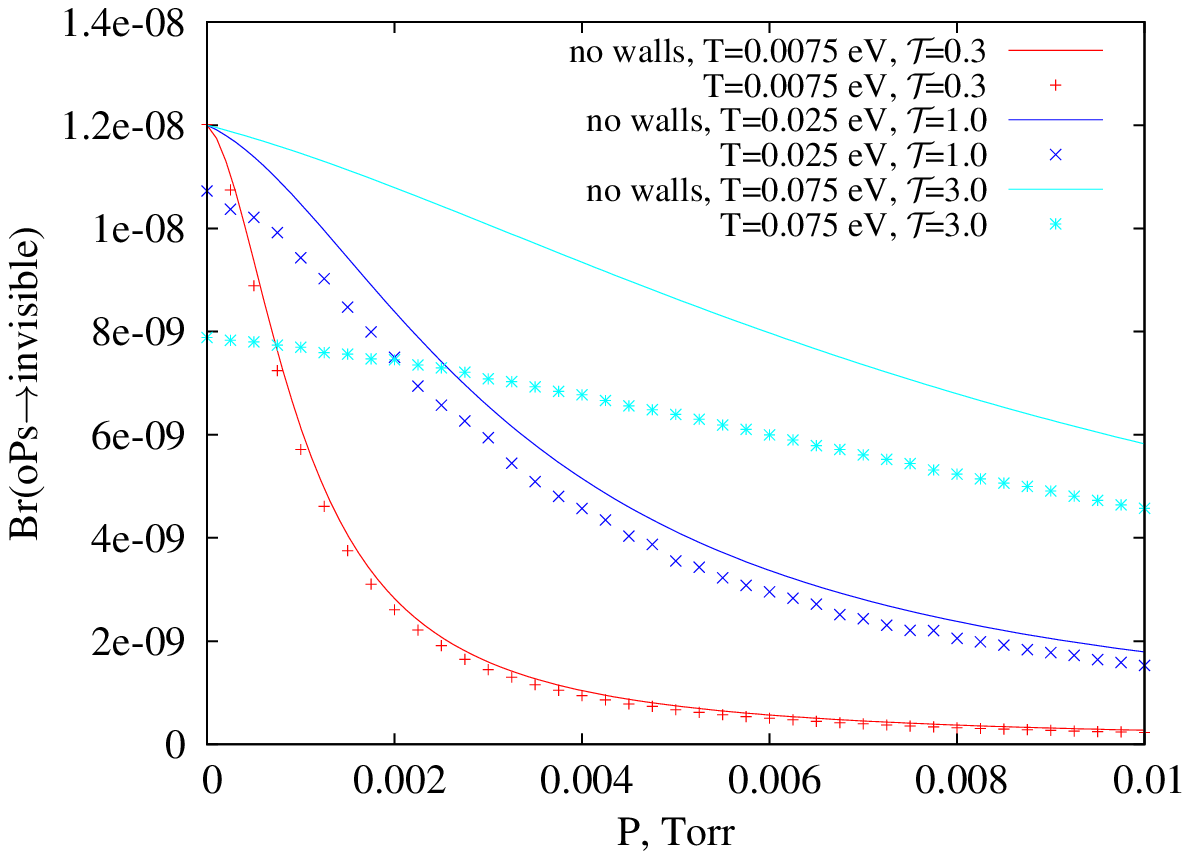}  
&
\includegraphics[angle=0,width=0.45\columnwidth]{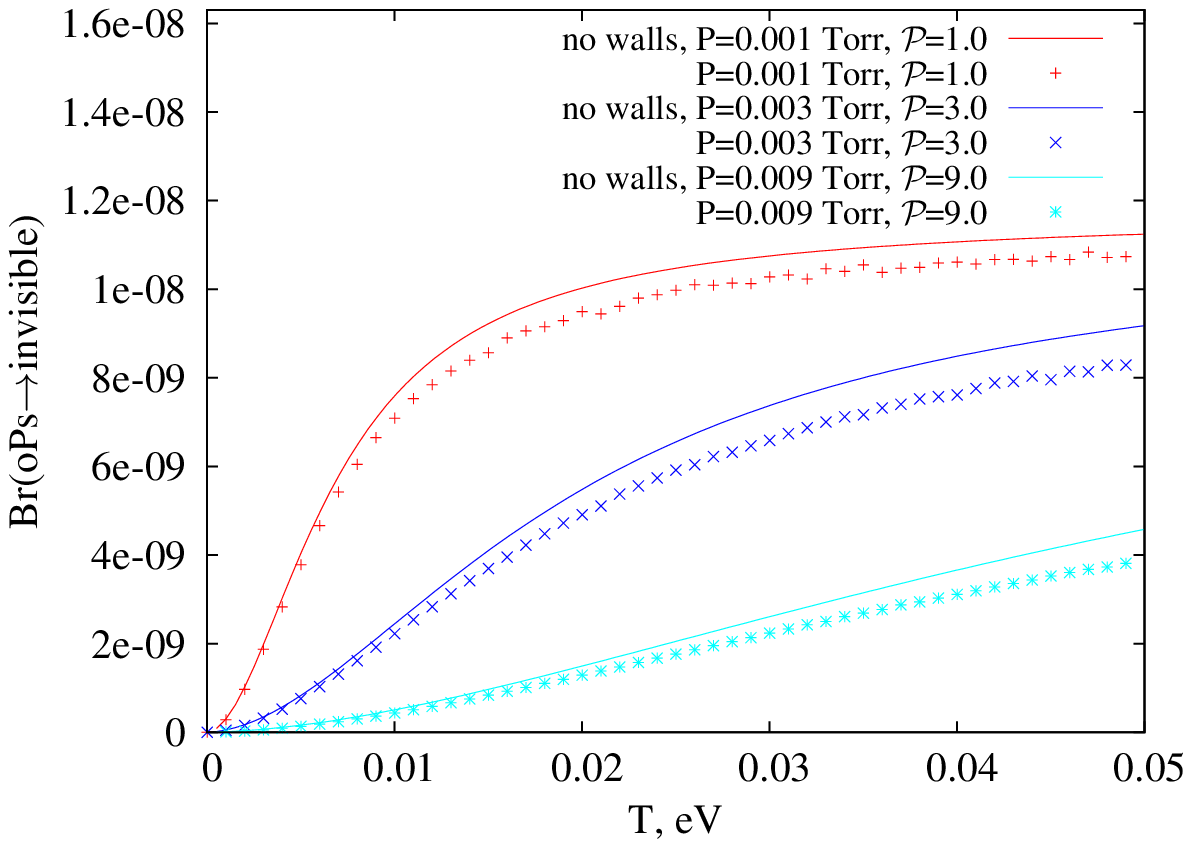}  
\end{tabular}
\caption{\label{press} The dependence of the branching ratio of
  the positronium disappearance on gas (here is nitrogen) pressure (left panel) and
  gas temperature (right panel) with (points) and without (solid
  lines, Eq.~\eqref{main-in-numbers}) wall collisions. The positronium
  temperature equals nitrogen temperature and we set 
  $\epsilon=10^{-9}$. Note that ${\rm Br}(\mathrm{oPs}\to
  \mathrm{invisible})$ scales as $\epsilon^2$.  
}
\end{figure}

The dependencies of the branching ratio of orthopositronium disappearance
on gas temperature and gas pressure are presented in Fig.~\ref{press}.
To see clearly the effect of wall collisions we also plot the pressure
dependencies obtained with Eq.~\eqref{main-in-numbers} for
comparison. Obviously, the cavity finite size effect as well as the gas
effect are always in
a reduction of positronium disappearance branching ratio: collisions
prevent positronium from oscillations to its twin.

\subsection{Oscillations in static electric and magnetic fields:
  Zeeman and Stark effects}
\label{Sec:fields}

It is well known that in an external static magnetic field $B$ the
state of orthopositronium with $m=0$ mixes with
parapositronium~\cite{Halpern:1954zz,Rich:1981nc}
and these states show Zeeman shifts in their energies and decay
rates\footnote{In the limit $B\to 0$ the state ``$+$'' goes to the
  orthopositronium state while ``$-$'' to parapositronium.}
\begin{equation}
\Delta_{\pm} =
\frac{1}{2}\l -\Delta_{HFS}\pm\sqrt{\Delta_{HFS}^2+(4\mu_{0}B)^2}\r, 
\end{equation}
\begin{equation}
\Gamma_{\pm} = \frac{1}{2}\l\gamma+\gamma_{P} \pm
\frac{\l \gamma-\gamma_{P} \r\,\Delta_{HFS}}{\sqrt{\Delta_{HFS}^2 
+ (4\mu_{0}B)^2}}\r,
\end{equation}
where $\Delta_{HFS}\approx 8.4\times 10^{-4}$~eV is hyperfine energy
splitting~ \cite{Sasaki:2010mg}, $\gamma_{P}\approx 8.1\times
10^{9}$~s$^{-1}$ is parapositronium decay rate and $\mu_{0}\approx
5.79\times 10^{-5}$~eV$/$T is the Bohr magneton. Here the conditions
$\gamma, \gamma_{P} \ll \Delta_{HFS}$ were used and we neglected
radiative corrections to gyromagnetic ratios of electron and
positron. 

If the magnetic field is strong, {\it i.e.} $4\mu_{0}B \gg 
\Delta_{HFS}$ which corresponds to $B\gg 3.6~$T, then
$\Gamma_{\pm} \approx \frac{1}{2}(\gamma+\gamma_{P}) \gg
\gamma$. Hence, the state  of orthopositronium with $m=0$ quickly
decays and its contribution to oPs-oPs$^{\prime}$ oscillations is
negligibly small. When
the magnetic field is weak, $4\mu_{0}B \ll  
\Delta_{HFS}$, the shifts in 
the energy levels and decay rates read 
\[
\Delta_{+} \approx \frac{(2\mu_{0}B)^2}{\Delta_{HFS}},\;\;
\Delta_{-} \approx -\Delta_{HFS} - \frac{(2\mu_{0}B)^2}{\Delta_{HFS}},
\]
\[
\Gamma_{+} \approx \left[
  1-\l\frac{2\mu_{0}B}{\Delta_{HFS}}\r^{\!\!2\;}\right]
\gamma  
+ \l\frac{2\mu_{0}B}{\Delta_{HFS}}\r^{\!\!2}\gamma_{P}\;,\;\;\;\;
\Gamma_{-} \approx \l\frac{2\mu_{0}B}{\Delta_{HFS}}\r^{\!\!2}\gamma
+ \left[ 1-\l\frac{2\mu_{0}B}{\Delta_{HFS}}\r^{\!\!2\;}\right]\gamma_{P}\;.
\]
Thus, when the weak magnetic field is applied the oPs state becomes a
superposition of ``$+$'' and ``$-$'' states of which ``$-$'' decays
quickly, while ``+'' is approximately $m=0$ state of the
orthopositronium which acquires shifts in its energy and decay rate. 
Numerically we estimate
\be
\Delta_{+}\approx 2.5\times 10^6~{\rm s}^{-1}\times \l\frac{B}{100~{\rm 
    G}}\r^2. 
\ee
The mixing between new state and the mirror twin of $m=0$ state of the 
orthopositronium also gets corrections of order
$(\mu_0 B/\Delta_{HFS})^2$ which can be neglected. 

There is also the quadratic Zeeman effect in
positronium~\cite{Feinberg:1989dq}. The corresponding energy shift
\begin{equation}
\Delta_{D} = 2\alpha^2 a_{0}^{3} B^2 \approx 10^4~{\rm s}^{-1}\times
\l \frac{B}{100~{\rm G}}\r^2\;, 
\end{equation}
where $a_0\simeq 0.1$\,nm is the Bohr radius, applies equally to all four
spin states of the lowest energy and can be considerable for large
magnetic field. 

In an external static electric field $E$ all three orthopositronium
states get shifted equally due to the Stark effect~\cite{landau3}, 
\be
\label{Stark-splitting}
\Delta_{S}=-\frac{1}{2}\alpha_0\epsilon_0 E^2=-1.85\times 10^{-7}\,{\rm
  eV}\times  \l \frac{E}{100\,{\rm kV/cm}}\r^2 \approx -2.8\cdot
  10^8~{\rm s}^{-1}\times \l \frac{E}{100\,{\rm kV/cm}}\r^2\;,
\ee
where $\alpha_0=\frac{9}{2}4\pi a_{0}^{3}$ determines polarizability of
orthopositronium,  and $\epsilon_0$ refers to the free space
  permittivity. With a reasonable field strength available in a laboratory (in
  particular, in the AEgIS experiment),  
$E\sim 100$\,kV/cm, the energy splitting due to the Stark
effect \eqref{Stark-splitting} can be considerable. 

Let us summarize the effect of the external electromagnetic field. For
large magnetic field, $m=0$ state immediately decays and does not
contribute to the oscillations. Hence, in case of unpolarized
orthopositroniums this results in decreasing of total oscillation
probability by a factor $2/3$. This behavior gives a tool to check for
possible systematics in the experiment, where magnetic fields as large
as 10\,T are available (as, in particular, in the
experiment~\cite{Crivelli:2010bk}), which is quite useful especially
if evidence for the oscillations is found.  In the weak magnetic
and electric fields the $m=0$ oPs state acquires total energy shift 
\be 
\Delta_{0} = \Delta_{+} + \Delta_{D} + \Delta_{S}\;, 
\ee 
and its decay width becomes $\gamma_{0}=\Gamma_{+}$. The
oPs states with $m=\pm 1$ get energy shift 
\be 
\Delta_{1} =\Delta_{D} + \Delta_{S}\;, 
\ee 
while their decay rates remain intact.

Averaging the disappearance probability over states with
different angular momentum components
\be
{\rm Br}\l \mathrm{oPs}\to
\mathrm{invisible}\r=
\frac{\int_0^{\infty}\rho_{2}^{m=0} dt+2\int_0^{\infty}\rho_{2}^{m=\pm 1} dt
}  {\int_0^{\infty}(\rho_{1}^{m=0}+2\rho_{1}^{m=\pm 1}+\rho_{2}^{m=0}
+2\rho_{2}^{m=\pm 1}) dt} \;,
\ee
one obtains 
\be
{\rm Br}\l \mathrm{oPs}\to
\mathrm{invisible}\r = \frac{\delta^2}{2\gamma(1 + 2\frac{\Gamma_{+}}{\gamma})} 
\l\frac{\Gamma_{0}}{\Delta_{0}^2 + \Gamma_{0}^{2}} +
2\frac{\Gamma_{+}}{\gamma}\frac{\Gamma}{\Delta_{1}^2+\Gamma^2}\r\;,
\ee
where $\Gamma_{0} = \frac{1}{2}(\Gamma_{+} + \gamma + w)$. 

\begin{figure}[!htb]
\includegraphics[angle=0,width=0.80\columnwidth]{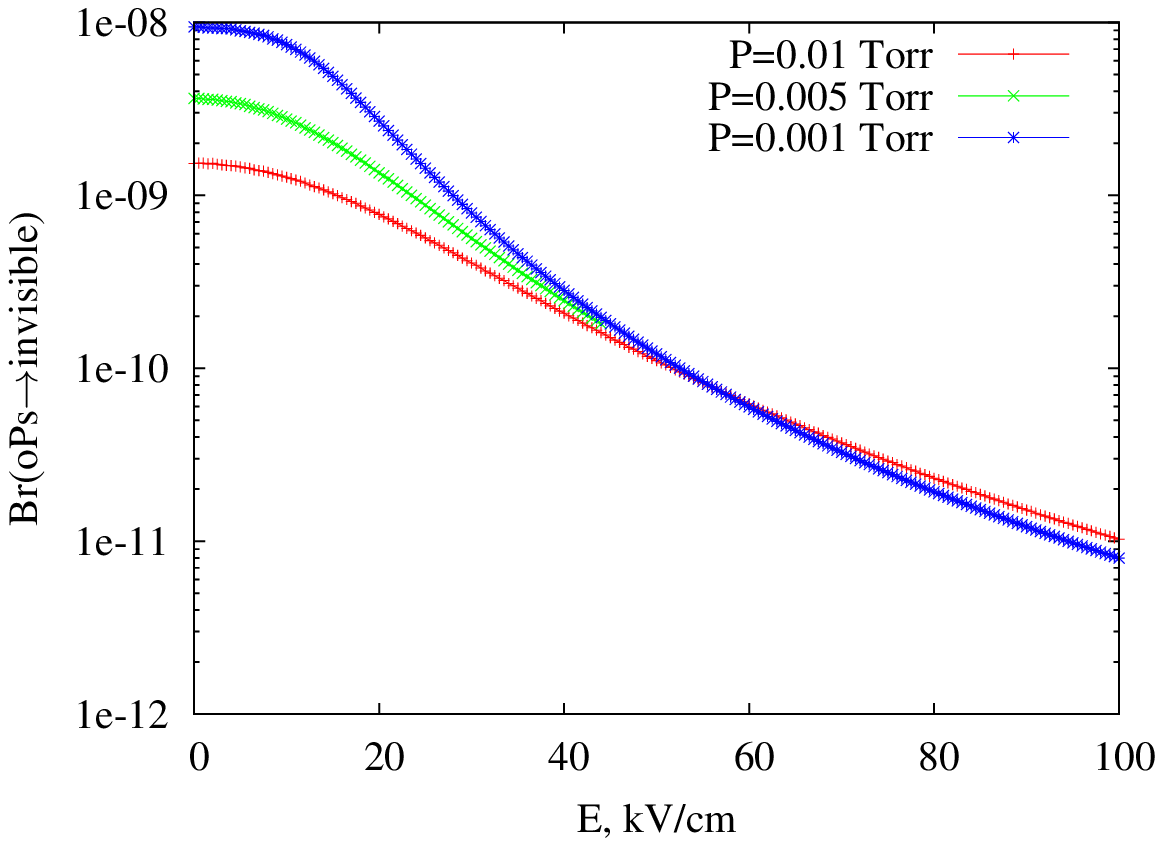} 
\caption{\label{E} The dependence of the branching ratio of the
  positronium disappearance on the external electric field $E$ at
  different values of gas pressure. We chose $\epsilon=10^{-9},
  k_{B}T_{\mathrm{oPs}}=0.025$~eV and the magnetic field $B=0$.
}
\end{figure}

\begin{figure}[!htb]
\includegraphics[angle=0,width=0.80\columnwidth]{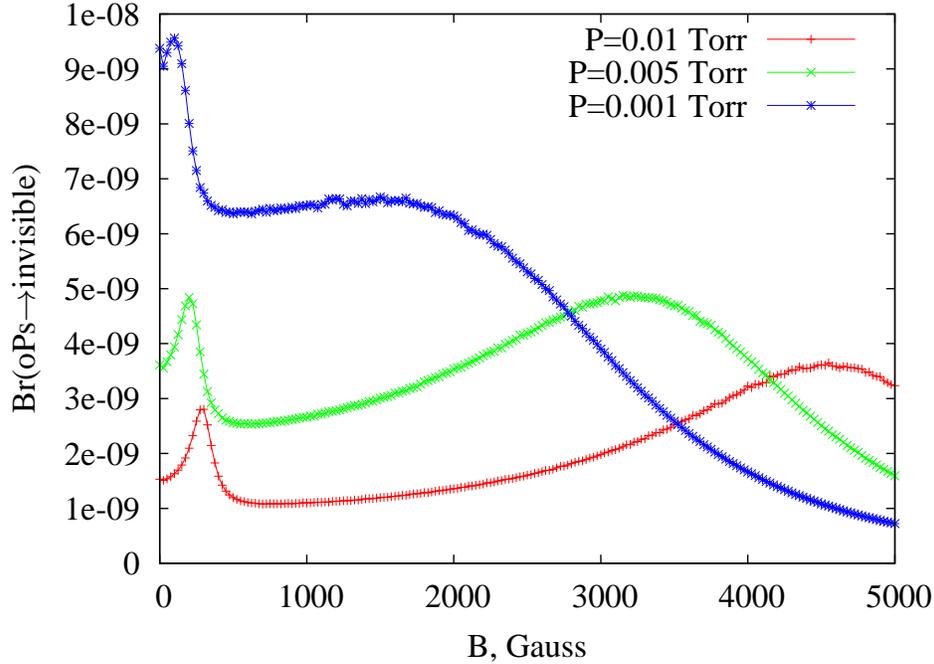} 
\caption{\label{B} The dependence of the branching ratio of the
  positronium disappearance on the magnetic field $B$ at different values
  of gas pressure. We chose $\epsilon=10^{-9}, k_{B}T_{\mathrm{oPs}}=0.025$\,eV
  and the electric field $E=0$. 
}
\end{figure}

The results for invisible branching ratio of the orthopositronium in
the presence of electric and magnetic fields with different
values of gas pressure are presented in
Figures~\ref{E} and~\ref{B}, respectively. 
Here we also include the effect of the wall collisions in the geometry
described in Section \ref{AEgIS}. 
We see that generically presence of an external electromagnetic field
leads to decreasing of the oscillation rate and thus to deterioration of
the sensitivity of experiment to the Mirror World physics. 
Some nontrivial dependence on the
magnetic field stems from the fact that both Zeeman effect contributions
to the energy splitting have the sign which is opposite to the
contribution of the coherent scattering~\eqref{hamiltonian-in-gas}. As
we have already mentioned the dependence on the external parameters like
electric or magnetic fields can be used to get rid of systematic
uncertainties or (in case of positive signal) to get evidence that the
observed effect is due to the oscillation nature of positronium
disappearance 
rather than direct decay into some invisible particles.

Let us note in passing, that in a model with broken mirror symmetry
where the positronium and its mirror twin obtain mass splitting larger
than orthopositronium width, oscillations are absent because of the
loss of coherence, and the  branching ratio of an invisible mode is
strongly suppressed (see Eq.~\eqref{main}),
\[
{\rm Br}\l \mathrm{oPs}\to \mathrm{invisible}\r\simeq \frac{\delta^2}{2}
\frac{\Gamma}{\gamma} \frac{1}{\Delta^2}\;. 
\] 
However, with magnetic field tuned to cancel this mass splitting by
using 
the Zeeman effect, one still has a possibility of the Mirror World
hunting via the positronium portal.

\section{Rydberg states and models with vacuum mass splitting between twins}
\label{Sec:Rydberg}

Let us proceed with discussion of oscillations of (highly) excited
positronium with zero total momentum $l=0$ and the large principal
quantum number $n$. Its annihilation rate into photons to the leading
order in QED coupling reads 
\be
\label{Rydberg-decay}
\Gamma_{\mathrm{oPs}_n\to3\gamma} = \gamma\,\frac{n+1}{2n^3}\;,  
\ee
thus numerically, 
\be
\label{Rydberg-decay-time}
\Gamma_{\mathrm{oPs}_n\to3\gamma}\simeq 7\times10^6\,{\rm
  s}^{-1}\times\frac{n+1}{2n^3}\;. 
\ee
At large $n$ it scales as $\propto n^{-2}$ and becomes much smaller
than the decay rate of the orthopositronium ground state. 
Meanwhile, the fastest transition from the level $ns$ 
is to the level $2p$, in which the rate is 
\begin{equation}
\label{Rydberg-transition}
\Gamma_{ns\to 2p}=\frac{4}{3}\left[\frac{m\alpha^2}{4}
\l\frac{1}{4}-\frac{1}{n^2}\r\right]^3\times |d|^2\;,
\end{equation} 
where $m$ is the electron mass, and the squared matrix element of the
dipole moment  
\cite{landau3} is given by
\[
|d|^2=\frac{4}{m^2\alpha}\frac{2^{15}n^9(n-2)^{2n-6}}{3(n+2)^{2n+6}}\;.
\]
At a  large $n$ it approaches 
\[
|d|^2\approx\frac{4}{m^2\alpha}\frac{2^{15}\,{\rm exp}(-8)}{3\,n^3}\;,
\]
and thus the transition rate \eqref{Rydberg-transition} asymptotes to 
 \be
\label{Rydberg-transition-asymptote}
\Gamma_{ns\to2p}\simeq\frac{2^7}{3^2\cdot {\rm exp}(8)}\frac{\alpha^5
  m}{n^3} \approx 7.5 \times 10^7\,{\rm s}^{-1}\times \frac{1}{n^3}\;.
\ee

The estimates in \eqref{Rydberg-decay-time} and
\eqref{Rydberg-transition-asymptote} explain what happens to the
Rydberg positroniums. Low excited levels of $n<20$ decay
into $2p$ state, which quickly ($\Gamma_{2p\to
  1s}=3.1\times10^8$\,s$^{-1}$)  decays further into $1s$, where
finally positronium 
annihilates. At a larger $n$ the direct annihilation of excited states
dominates. Note that the matrix element of the positronium oscillation 
\eqref{positronium-mixing} also depends on $n$. The estimate of the
total oscillation probability of the excited positronium 
follows from the formula in \eqref{branching-in-vacuum} upon 
rescaling 
\[
\delta\to\delta \,\frac{n+1}{2\,n^3} \;,~~~~
\gamma \to \Gamma_{\mathrm{oPs}_n\to3\gamma}=\gamma\,\frac{n+1}{2\,n^3}\;. 
\]
Hence, at large $n$ the invisible decay branching ratio of the Rydberg
positronium coincides with that of the ground state.  

Note that for the  model with small vacuum splitting between
positronium and its twin one can think of oscillations between states
of different principal numbers $n$ and $n'$. Then both $E$ and $E'$,
and $\gamma$ in two diagonal entries depend on these numbers 
and the formula for the disappearance 
branching ratio can be generalized to this case.

\section{Conclusions and open problems}
\label{Sec:Conclusions}

In this paper we have presented the complete analysis of orthopositronium
oscillations into its twin within the Mirror World models. We took into
account the relevant effects due to the possible scatterings off walls and gas
atoms, and the influence of external electric and magnetic fields. 

In a background-free case, the highest sensitivity to the mixing of a
photon and paraphoton responsible for the oscillations in positronium
sector, is achieved in the pure vacuum with an infinitely large experimental
volume. At a given value of mixing parameter the invisible branching ratio of
positronium (disappearing via oscillation to its twin, which
subsequently decays into paraphotons) decreases when applying electric
or magnetic fields, adding gas to the cavity and decreasing its size
(or equivalently increasing positronium velocities). We have calculated
this decrease as a function of the relevant physical parameters, which can
be used to get rid of possible systematics in the experiment, if 
evidence of a positronium disappearance is found. 

To perform numerical estimates, a computer code has been written,
which allows us to account for a realistic geometry of the cavity,
a realistic positronium spectrum (distribution over velocity), the details
of scattering off the cavity walls and gas atoms, etc. To illustrate
the code, the numerical results for the setup of positronium
experiment~\cite{Crivelli:2010bk} proposed within the AEgIS project have
been presented. For the first time, the experiment will enter the
phenomenologically interesting and cosmologically allowed region of the
model parameter space.  The code can be used for other setups and may
be further improved by implementing more details of positronium
interactions with gas, walls, and magnetic and electric fields. The
properly completed modification can be used in simulations of events in a
real experiment. 

The presented numerical results for the experimental
setup~\cite{Crivelli:2010bk} can be improved with the account of
high-velocity  
positroniums. Their presence in a small number seems to slightly
diminish the sensitivity to the mixing parameter: faster 
scatterings off gas atoms and walls spoil oscillations to the Mirror
World. A similar effect --- decreasing sensitivity to invisible mode ---
is expected for a more accurate realistic treatment of positronium
scatterings including induced positronium annihilation, 
small blind spots in the detector, etc. Thus, we propose our
numerical estimates to be used to place an upper limit on the
sensitivity of a given experiment in the mixing parameter.    

For the first time, we analyzed oscillations of excited positroniums
--- Rydberg positroniums --- into their similarly excited twins. In the
vacuum the branching ratio of a high-level Rydberg positronium decay
into nothing is the same as that of the ground state (at the same
mixing). Rydberg positroniums will be available, {\it e.g.}, in the AEgIS
experiment and can be used for better control over possible
systematics. They might 
be of some interest in the models with a slightly violated Mirror
symmetry, resulting in a small mass shift between electron and its
twin. 

We thank P. Crivelli for correspondence and S. Gninenko for numerous
valuable discussions.  The work is supported in part by the grant of
the President of the Russian Federation NS-5525.2010.2, by Russian
Foundation for Basic Research grants 11-02-01528-a (D.G. and S.D.) and
11-02-92108-YAF\_a (D.G.), by the SCOPES grant (D.G.) and by the
"Dynasty" Foundation awarded by the Scientific board of ICPFM (A.T.).


\end{document}